# DCT and Eigenvectors of Covariance of 1st and 2nd order Discrete fractional Brownian motion

Anubha Gupta[1] and ShivDutt Joshi[2]
[1]Signal Processing and Communication Research Center, IIIT-Hyderabad, India
[2]Electrical Engineering Department, Indian Institute of Technology-Delhi, India

*Abstract*— **This paper establishes connection between discrete cosine transform (DCT) and 1st and 2nd order discrete-time fractional Brownian motion process. It is proved that the eigenvectors of the auto-covariance matrix of a 1st and 2nd order discrete-time fractional Brownian motion can be approximated by DCT basis vectors in the asymptotic sense. Perturbation in eigenvectors from DCT basis vectors is modeled using the analytic perturbation theory of linear operators.**

## I. INTRODUCTION

Fractional Brownian motion (fBm) processes are Gaussian non-stationary random processes that form a class of statistically self-similar processes [1]. In the past few decades, these processes have been utilized in many applications in the engineering fields, including characterization of textures in bone radiographs, image processing and segmentation, medical image analysis [2]-[4] and network traffic analysis [5], [6], etc. The statistical properties of an fBm process are characterized by a single parameter called Hurst exponent $H$. In addition, the process is statistically self-similar and has stationary increments. That is why an fBm process is also called an $H$-sssi process (or $H$-self similar with stationary increments) [1].

Because, in general, we work with the discrete-time processes, the covariance structure of 1st-order discrete-time fractional Brownian motion (dfBm) was explored in [7]. Following were the salient points of [7]: 1) The covariance structure of 1st order discrete-time fractional Brownian motion (dfBm) was explored. 2) The work focused on the modeling of eigenvalues. It was shown that the eigenvalues of the covariance matrix are dependent on Hurst exponent characterizing the discrete-time fractional Brownian motion. Only one eigenvalue depends on time index *n*, while others are time-invariant in the asymptotic sense. 3) The covariance matrix is diagonalizable by a time-invariant matrix in the asymptotic sense. Thus, asymptotically, the eigenvectors are time- invariant. Although in [7], it was observed that the eigenvectors of the covariance of a 1$^{st}$ order dfBm can be used as filters varying from lowpass to highpass filters, these eigenvectors were not studied further.

In this paper, we study the eigenvectors of 1$^{st}$ and 2$^{nd}$ order dfBm. This work will be of significance in applications such as subband decomposition, where we would like to work with the eigenvectors of a dfBm process. For example, [8] utilizes the filterbank structure constructed from the eigenvectors of the covariance matrix of 2$^{nd}$ order dfBm to denoise ECG signals. In such applications, results of this work will simplify processing with no compromise on the outcome of the application.

The outline of the paper is as follows: In Section 2, we briefly present some relevant results on the theory of fBm processes, toeplitz matrices, perturbation theory of linear operators, and nonstationary Gauss-Markov processes. In Section 3, we establish a connection between discrete cosine transform (DCT) and the diagonalizing unitary transform of the covariance matrix of 1$^{st}$ and higher order dfBm. Simulation results are presented in Section 4. Conclusions are presented in Section 5.

*Notations:* We use lowercase bold letters and uppercase bold letters to represent vectors and matrices, respectively. The scalar variables are represented by lowercase italicized letters. In addition, $E\{\cdot\}$ denotes the expectation operator.

## II. PRELIMINARIES

### A. Overview of fractional Brownian motion

A continuous-time random process is called self-similar if its statistical properties are scale invariant. Symbolically, it is represented as

$$x(ct) \stackrel{d}{=} c^H x(t), \qquad (1)$$

where random process $x(t)$ is self similar with self similarity index $H$ (Hurst exponent) for any scale parameter $c > 0$. This is to note that equality in (1) holds in statistical sense for all finite distributions [1].

An important class of these non-stationary self-similar processes is those with self-similarity index $H$ and having stationary increments ($H$-sssi), i.e.,

$$x(t+\tau) - x(\tau) \stackrel{d}{=} x(t) - x(0) \quad \text{for all } \tau \in \Re. \tag{2}$$

If the increments of a non-stationary, self-similar process with $0 < H < 1$ are stationary ($H$-sssi) and arise from a Gaussian distribution, the process is known as fractional Brownian motion (fBm) [1].

*$m^{th}$ order fractional Brownian motion:* Fractional Brownian motion with $0 < H < 1$ is called $1^{st}$ order fBm or 1-fBm, while an fBm process with Hurst exponent $H \in (m-1, m)$ is called $m^{th}$ order fBm or $m$-fBm and is obtained by integration of $(m-1)^{th}$ order fBm as below [9]:

$$B_H^m(t) = \int_0^t B_{H-1}^{m-1}(u)\,du \tag{3}$$

Corresponding to the discrete data set, $m^{th}$ order discrete-time fractional Brownian motion ($m$-dfBm) [9] is defined as

$$B_H^m(n) = B_H^m(nT_s), \tag{4}$$

where $T_s$ is the sampling period. Because the process is self-similar for any value of $c > 0$, therefore, $T_s$ can be taken to be equal to one without loss of generality. The mean value, variance, and the auto-covariance sequence of a 1-dfBm process, $B_H^1(n)$, are given as below [9]:

$$E\{B_H^1(n)\} = 0, \tag{5}$$

$$Var\{B_H^1(n)\} = n^{2H}\sigma_H^2, \tag{6}$$

$$r_{B,1}^H(n_1, n_2) = \frac{\sigma_H^2}{2}\left(|n_1|^{2H} - |n_1 - n_2|^{2H} + |n_2|^{2H}\right), \tag{7}$$

where $\sigma_H^2 = Var\{B_H^1(1)\} = \dfrac{1}{\Gamma(2H+1)|\sin(\pi H)|}$. Thus, a 1-dfBm process is a zero mean, self similar, non-stationary random process.

Using (3) and (7), it is easy to obtain the auto-covariance sequence of an $m$-dfBm process given as below:

$$r_{B,m}^H(n_1, n_2) = (-1)^m \frac{\sigma_H^2}{2} \left( |n_1 - n_2|^{2H} - \sum_{j=0}^{m-1} (-1)^j \binom{2H}{j} \left[ \left(\frac{n_1}{n_2}\right)^j |n_2|^{2H} + \left(\frac{n_2}{n_1}\right)^j |n_1|^{2H} \right] \right), \tag{8}$$

where $\binom{2H}{j} = \frac{2H(2H-1)\cdots(2H-j+1)}{j!}$

and for $m \geq 2$  $\quad \sigma_H^2 = \frac{\sigma_{H-m+1}^2}{(2H)(2H-1)\cdots(2H-(2m-3))} = \frac{1}{\Gamma(2H+1)|\sin(\pi H)|}.$

### B. Brief Review of Prior Work on the Structure of Covariance Matrix of 1-dfBm Process

Consider a length $M$ vector, $\mathbf{x}(n)$, of random variables of a zero mean 1-dfBm process

$$\mathbf{x}(n) = \begin{bmatrix} B_H^1(n) & B_H^1(n+1) & \cdots & B_H^1(n+M-1) \end{bmatrix}^T. \tag{9}$$

The auto-covariance of $\mathbf{x}(n)$ is given as below:

$$\mathbf{R}_{B,1}^H(n) = E\{\mathbf{x}(n)\mathbf{x}^T(n)\} \tag{10}$$

$$\mathbf{R}_{B,1}^H(n) = E\left( \begin{bmatrix} B_H^1(n) \\ B_H^1(n+1) \\ \vdots \\ B_H^1(n+M-1) \end{bmatrix} \begin{bmatrix} B_H^1(n) & B_H^1(n+1) & \cdots & B_H^1(n+M-1) \end{bmatrix} \right). \tag{11}$$

On using (7) in (11), we obtain

$$\mathbf{R}_{B,1}^H(n) = \frac{\sigma_H^2}{2} \begin{bmatrix} 2n^{2H} & n^{2H} + (n+1)^{2H} - 1 & n^{2H} + (n+2)^{2H} - 2^{2H} & \cdots & n^{2H} + (n+M-1)^{2H} - (M-1)^{2H} \\ n^{2H} + (n+1)^{2H} - 1 & 2(n+1)^{2H} & \cdot & \cdots & \cdot \\ \vdots & \vdots & \vdots & \cdots & \vdots \\ n^{2H} + (n+M-1)^{2H} - (M-1)^{2H} & \cdot & \cdot & \cdots & 2(n+M-1)^{2H} \end{bmatrix}. \tag{12}$$

The auto-covariance matrix, $\mathbf{R}_{B,1}^H(n)$, of a random vector $\mathbf{x}(n)$ (as defined in (9)) for any time index $n$ is a real symmetric matrix and hence, is a diagonalizable matrix. In [7], the structure of the auto-covariance matrix of a 1st order discrete-time fBm (1-dfBm) process was studied. Two theorems, from [7], relevant for this work are reproduced below.

*Theorem-2.1:* The auto-covariance matrix $\mathbf{R}_{B,1}^{H}(n)$ of a vector $\mathbf{x}(n)$ of random variables of length $M$ of a discrete-time 1st order fractional Brownian motion with $0<H<1$ can be approximated as (symbol '^' is used to denote approximation)

$$\hat{\mathbf{R}}_{B,1}^{H}(n) = \frac{\sigma_H^2}{2} 2n^{2H} (\mathbf{A} + \varepsilon_1 \mathbf{B} + \varepsilon_2 \mathbf{C}), \tag{13}$$

where

$$\mathbf{A} = \begin{bmatrix} 1 & 1 & \cdots & 1 \\ \cdot & \cdot & \cdots & \cdot \\ \cdot & \cdot & 1 & \cdots & \cdot \\ \cdot & \cdot & \cdot & 1 & \cdot & \cdot \\ \cdot & \cdot & \cdot & \cdot & \cdot & \cdot \\ 1 & \cdots & & & & 1 \end{bmatrix}, \quad \mathbf{B} = \begin{bmatrix} 0 & 1 & 2 & \cdots & (M-1) \\ 1 & 2 & 3 & \cdots & M \\ 2 & 3 & \cdots & & (M+1) \\ \cdot & \cdot & \cdot & & \cdot \\ (M-2) & \cdots & & & (2M-3) \\ (M-1) & \cdots & & & (2M-2) \end{bmatrix}, \quad \mathbf{C} = -\begin{bmatrix} 0 & 1 & 2^{2H} & \cdots & (M-1)^{2H} \\ 1 & 0 & 1 & \cdots & (M-2)^{2H} \\ \cdot & \cdot & \cdot & \cdots & \cdot \\ \cdot & \cdot & \cdot & \cdots & 1 \\ (M-1)^{2H} & \cdot & \cdot & \cdots & 0 \end{bmatrix},$$

$\varepsilon_1 = \frac{H}{n}$, and $\varepsilon_2 = \frac{1}{2n^{2H}}$ for $n > n_{\min}$.

The normalized approximation error $e(n)$ defined as the Frobenius norm of the difference between the true $\mathbf{R}_{B,1}^{H}(n)$ and its approximation $\hat{\mathbf{R}}_{B,1}^{H}(n)$ is bounded from above as

$$e(n) \triangleq \frac{\|\mathbf{R}_{B,1}^{H}(n) - \hat{\mathbf{R}}_{B,1}^{H}(n)\|_F}{\|\mathbf{R}_{B,1}^{H}(n)\|_F} \leq |\alpha|, \tag{14}$$

beyond $n > n_{\min}$ with $n_{\min} = \left\lceil (M-1)H \cdot \left(\left|2 - \frac{1}{H}\right| \cdot \frac{1}{|\alpha|}\right)^{\frac{1}{2}} \right\rceil$.

*Theorem 2.2 (Structure Theorem):* The $M \times M$ auto-covariance matrix $\mathbf{R}_{B,1}^{H}(n)$ of a discrete-time 1st order fractional Brownian motion with $0<H<1$ can be approximated as $\tilde{\mathbf{R}}_{B,1}^{H}(n)$ for large $n$, such that

$$\tilde{\mathbf{R}}_{B,1}^{H}(n) = \mathbf{Q} \tilde{\mathbf{\Lambda}}_{B,1}^{H}(n) \mathbf{Q}^{\mathrm{T}}, \tag{15}$$

where $\mathbf{Q}(n) \approx \mathbf{Q}$ is a constant orthogonal matrix in the asymptotic sense and $\tilde{\mathbf{\Lambda}}_{B,1}^{H}(n) = \mathrm{diag}\{\lambda_1, \lambda_2, ..., \lambda_M(n)\}$ for large $n$.

From Theorem 2.2, it is evident that all the eigenvalues of $\mathbf{R}_{B,1}^{H}(n)$ except for one are time-invariant for large $n$ [7]. The largest eigenvalue is modeled as a function of the time index $n$ and the Hurst exponent $H$ in [7]. In [7], the structure of the orthogonal matrix $\mathbf{Q}$ was not explored which is the goal of this paper.

In this paper, we explore the structure of the orthogonal matrix $\mathbf{Q}$ analytically for 1$^{st}$ order and 2$^{nd}$ order dfBm processes. To this end, we require some results from the analytic perturbation theory of linear operators, a brief review of which is presented below.

*C. Results from the perturbation theory of linear operators*

In this section, we present some relevant results from the matrix theory on analytic perturbation of linear operators [10].

Let $\mathbf{A} \in \mathbb{C}^{M \times M}$ be an unperturbed matrix corresponding to a linear operator and suppose that $\lambda_1, \lambda_2, \ldots \lambda_s$ are the distinct eigenvalues of $\mathbf{A}$, so that

$$p(\lambda) = (\lambda - \lambda_1)^{m_1}(\lambda - \lambda_2)^{m_2} \ldots (\lambda - \lambda_s)^{m_s}. \tag{16}$$

is the minimal polynomial of $\mathbf{A}$ with $m_i$ being the multiplicity of eigenvalue $\lambda_i$. If a given function $f(\lambda)$ is defined on the spectrum $\sigma(\mathbf{A}) = \{\lambda_1, \lambda_2, \ldots \lambda_M\}$ of $\mathbf{A}$, i.e., the value of $f(\lambda)$ at the eigenvalue $\lambda_k$ ($k=1,2,\ldots s$, $r=0,1,\ldots m_k-1$) denoted as $f_{k,r}$, exists, then there exist component matrices $\mathbf{Z}_{kr}$'s independent of $f(\lambda)$ such that

$$f(\mathbf{A}) = \sum_{k=1}^{s} \sum_{r=0}^{m_k-1} f_{k,r} \mathbf{Z}_{kr} \tag{17}$$

The matrices $\mathbf{Z}_{kr}$ are linearly independent, belong to $\mathbb{C}^{M \times M}$, and commute with $\mathbf{A}$ and with each other. These component matrices $\mathbf{Z}_{kr}$ ($k=1,2,\ldots s$, $r=0,1,\ldots m_k-1$) satisfy the following conditions:

$$\sum_{k=1}^{s} \mathbf{Z}_{k0} = \mathbf{I}; \tag{18a}$$

$$\mathbf{Z}_{kp}\mathbf{Z}_{lm} = 0 \quad \text{if} \quad k \neq l; \tag{18b}$$

$$\mathbf{Z}_{kr}^2 = \mathbf{Z}_{kr} \quad \text{iff} \quad r = 0; \tag{18c}$$

$$\mathbf{Z}_{kr}\mathbf{Z}_{k0} = \mathbf{Z}_{kr}, r = 0,1,....m_k - 1. \tag{18d}$$

The matrices $\mathbf{Z}_{k0}$ for $k=1,2,.....s$ are called projectors and their sum in (18a) above is known as the resolution of identity. Further, if these components $\mathbf{Z}_{k0}$ for $k=1,2,.....s$ are known, then we can find all the component matrices using (19) as

$$\mathbf{Z}_{kr} = \frac{1}{r!}(\mathbf{A} - \lambda_k \mathbf{I})^r \mathbf{Z}_{k0}. \tag{19}$$

The resolvent $\mathbf{R}_z = (z\mathbf{I} - \mathbf{A})^{-1}$ of $\mathbf{A}$ defined on $z \notin \sigma(\mathbf{A})$ can be expressed in terms of the component matrices of $\mathbf{A}$ as

$$\mathbf{R}_z = (z\mathbf{I} - \mathbf{A})^{-1} = \sum_{k=1}^{s}\sum_{r=0}^{m_k-1}\frac{r!}{(z-\lambda_k)^{r+1}}\mathbf{Z}_{kr}. \tag{20}$$

Suppose that $\mathbf{A}(\varepsilon)$ denotes the perturbation of a matrix $\mathbf{A}$. The elements of $\mathbf{A}(\varepsilon)$ are analytic functions of $\varepsilon$ in a neighborhood of $\varepsilon = 0$ such that

$$\mathbf{A}(\varepsilon) = \mathbf{A} + \varepsilon \mathbf{A}^{(1)} + \varepsilon^2 \mathbf{A}^{(2)} + .......... \tag{21}$$

where $\varepsilon$ is a complex number.

Because the eigenvalues $\lambda_1(\varepsilon), \lambda_2(\varepsilon), ..... \lambda_M(\varepsilon)$ of $\mathbf{A}(\varepsilon)$ depend continuously on $\varepsilon$, we may suppose that $\lambda_j(\varepsilon) \to \lambda_j$ as $|\varepsilon| \to 0$ for $j=1,2,....M$. Then, it is known that [10]

(a) If $\lambda_j$ is an unrepeated eigenvalue of $\mathbf{A}$, then $\lambda_j(\varepsilon)$ is analytic in a neighborhood of $\varepsilon = 0$;

(b) If $\lambda_j$ has algebraic multiplicity $m$ and $\lambda_{jr}(\varepsilon) \to \lambda_j$ for $r=1,2....m$, then $\lambda_{jr}(\varepsilon)$ is an analytic function of $\varepsilon^{1/l}$ in a neighborhood of $\varepsilon = 0$, where $l \leq m$ and $\varepsilon^{1/l}$ is one of the $l$ branches of the function $\varepsilon^{1/l}$.

For analytic and Hermitian matrices, the attention is confined to real $\varepsilon$ and it is assumed that the real and imaginary components of elements of the perturbed matrix $\mathbf{A}(\varepsilon)$ are real analytic functions of the real parameter $\varepsilon$. Under these assumptions, the perturbed eigenvalues remain real, the component matrices of $\mathbf{A}(\varepsilon)$ associated with perturbed eigenvalues will be Hermitian and, the perturbed eigenvalues are analytic in $\varepsilon$, whatever the multiplicity of the unperturbed eigenvalues may be. The perturbed component matrices are also analytic and hence, the eigenvectors of $\mathbf{A}(\varepsilon)$ are analytic and orthonormal throughout a neighborhood of $\varepsilon=0$.

*Result 2.1 [10] (Unrepeated eigenvalue):* Let $\mathbf{A}(\varepsilon)$ be a matrix that is analytic in $\varepsilon$ on a neighborhood of $\varepsilon = 0$, and suppose $\mathbf{A}(0)=\mathbf{A}$. Let $\lambda_j$ be an unrepeated eigenvalue of $\mathbf{A}$ with index (or degree) one, then, for sufficiently small $|\varepsilon|$, there is an eigenvalue $\lambda_j(\varepsilon)$ of $\mathbf{A}(\varepsilon)$ such that

$$\lambda_j(\varepsilon) = \lambda_j + \varepsilon \lambda_j^{(1)} + \varepsilon^2 \lambda_j^{(2)} + \ldots\ldots\ldots \tag{22}$$

Also, there are left and right eigenvectors $\mathbf{\psi}_j(\varepsilon)$ and $\mathbf{\varphi}_j(\varepsilon)$, respectively, associated with $\lambda_j(\varepsilon)$ for which

$$\mathbf{\psi}_j^{\mathrm{T}}(\varepsilon)\mathbf{\varphi}_j(\varepsilon) = 1 \tag{23}$$

and
$$\mathbf{\varphi}_j(\varepsilon) = \mathbf{\varphi}_j + \varepsilon \mathbf{\varphi}_j^{(1)} + \varepsilon^2 \mathbf{\varphi}_j^{(2)} + \ldots, \tag{24}$$

$$\mathbf{\psi}_j(\varepsilon) = \mathbf{\psi}_j + \varepsilon \mathbf{\psi}_j^{(1)} + \varepsilon^2 \mathbf{\psi}_j^{(2)} + \ldots \tag{25}$$

The complete solution for the first order perturbation coefficients is

$$\lambda_j^{(1)} = \mathbf{\psi}_j^{\mathrm{T}} \mathbf{A}^{(1)} \mathbf{\varphi}_j, \tag{26}$$

and $\mathbf{\varphi}_j^{(1)} = \mathbf{E}_j \mathbf{A}^{(1)} \mathbf{\varphi}_j,$ \tag{27}

where the matrices $\mathbf{E}_j$ are defined as

$$\mathbf{E}_j = \lim_{z \to \lambda_j}(z\mathbf{I} - \mathbf{A})^{-1}(\mathbf{I} - \mathbf{Z}_{j0}) = \sum_{\substack{k=1 \\ k \neq j}}^{s} \sum_{r=0}^{m_k - 1} \frac{r!}{(\lambda_j - \lambda_k)^{r+1}} \mathbf{Z}_{kr}. \tag{28}$$

If the perturbation of $\mathbf{A}$ is linear in $\varepsilon$ with $\mathbf{A} = \mathbf{A} + \varepsilon \mathbf{A}^{(1)}$, then the perturbation coefficients for all orders are given by

$$\lambda_j^{(k)} = \mathbf{\psi}_j^{\mathrm{T}} \mathbf{A}^{(1)} \mathbf{\varphi}_j^{(k-1)}, \tag{29}$$

$$\mathbf{\varphi}_j^{(k)} = \mathbf{E}_j\left(\mathbf{A}^{(1)}\mathbf{\varphi}_j^{(k-1)} - \sum_{p=1}^{k-1}\lambda_j^{(p)}\mathbf{\varphi}_j^{(k-p)}\right), \qquad \text{for } k=1,2,\ldots. \tag{30}$$

*Result 2.2 [10] (Repeated eigenvalue):* Let $\mathbf{A}(\varepsilon)$ be a matrix that is analytic in $\varepsilon$ on a neighborhood of $\varepsilon=0$, and suppose $\mathbf{A}(0) = \mathbf{A}$. Let $\lambda_j$ be an eigenvalue of $\mathbf{A}$ of index one and multiplicity $m$, then, the eigenvalue $\lambda_j$ splits into eigenvalues $\lambda_{j1}(\varepsilon)$, $\lambda_{j2}(\varepsilon)$,….. $\lambda_{jm}(\varepsilon)$ for sufficiently small $|\varepsilon|$. Let $\lambda_{jr}(\varepsilon)$ ($r = 1,2,……m$) be such an eigenvalue of $\mathbf{A}(\varepsilon)$ for which $\lambda_{jr}(0)= \lambda_j$. Then, there is a number $a_{jr}$ and a positive integer $l \leq m$ such that

$$\lambda_{jr}(\varepsilon) = \lambda_j + a_{jr}\varepsilon + 0(|\varepsilon|^{1+(1/l)}) \tag{31}$$

as $|\varepsilon| \to 0$ and $a_{jr}$ is an eigenvalue of $\mathbf{Z}_{j0}\mathbf{A}^{(1)}\mathbf{Z}_{j0}$,

where
$$\mathbf{Z}_{j0} = \frac{1}{2\pi i}\int_{L_j} \mathbf{R}_z dz \tag{32}$$

with $L_j$ being a simple closed contour that encircles $\lambda_j$, $\lambda_{j1}(\varepsilon)$, $\lambda_{j2}(\varepsilon)$,….. $\lambda_{jm}(\varepsilon)$ and no other eigenvalue of $\mathbf{A}$ or $\mathbf{A}(\varepsilon)$.

Corresponding to each $\lambda_{jr}(\varepsilon)$, there is an eigenvector $\boldsymbol{\varphi}_{jr}(\varepsilon)$ such that, for small enough $|\varepsilon|$

$$\boldsymbol{\varphi}_{jr}(\varepsilon) = \boldsymbol{\varphi}_r + \varepsilon^{1/l}\boldsymbol{\varphi}_{j,1} + \varepsilon^{2/l}\boldsymbol{\varphi}_{j,2} + ....., \tag{33}$$

where $\boldsymbol{\varphi}_r$, $\boldsymbol{\varphi}_{j,1}$, ….., $\boldsymbol{\varphi}_{j,l-1} \in \text{Ker}(\mathbf{A}-\lambda_j\mathbf{I})$ and $a_{jr}\boldsymbol{\varphi}_r = \mathbf{Z}_{j0}\mathbf{A}^{(1)}\mathbf{Z}_{j0}\boldsymbol{\varphi}_r$.

### D. DCT and 1-D Gauss-Markov Random Processes

Consider a length $M$ vector $\mathbf{x} = [x(1)\ x(2)\ .\ .\ .\ x(M)]^T$ of a zero-mean non-causal $1^{st}$ order Gauss Markov random (GMr) process defined by

$$x(n) = \alpha[x(n-1) + x(n+1)] + v(n) \quad 2 \le n \le M-1 \tag{34a}$$

$$(1 - k_1\alpha)x(1) = \alpha x(2) + v(2) \tag{34b}$$

$$(1 - k_2\alpha)x(M) = \alpha x(M-1) + v(M) \tag{34c}$$

where
$$\mathbf{v} = [v(1)\ v(2)\ .\ .\ .\ v(M)]^T \tag{35}$$

is a vector of a zero mean correlated noise process with covariance matrix

$$\mathbf{R}_\mathbf{v} = E\{\mathbf{v}\mathbf{v}^T\} = \beta^2 \mathbf{J}_D \tag{36}$$

and $\mathbf{J}_D = \mathbf{J}_D(k_1, k_2, 0, 0)$ is a tri-diagonal Jacobi matrix defined as

$$\mathbf{J}_D = \mathbf{J}_D(k_1, k_2, k_3, k_4) = \begin{bmatrix} 1-k_1\alpha & -\alpha & 0 & . & . & 0 & k_3\alpha \\ -\alpha & 1 & -\alpha & 0 & . & . & 0 \\ 0 & -\alpha & 1 & -\alpha & 0 & . & . \\ 0 & 0 & . & . & . & . & . \\ . & . & . & . & 1 & -\alpha & 0 \\ 0 & . & . & . & -\alpha & 1 & -\alpha \\ k_4\alpha & 0 & . & 0 & 0 & -\alpha & 1-k_2\alpha \end{bmatrix}. \tag{37}$$

Using (34) and (37), the vector $\mathbf{x}$ can also be written as

$$\mathbf{J}_D \mathbf{x} = \mathbf{v}. \tag{38}$$

From (36) and (38), it can be easily verified that

$$\mathbf{R}_{\mathbf{x}} = E\{\mathbf{x}\mathbf{x}^T\} = \beta^2 \mathbf{J}_D^{-1} \tag{39a}$$

or

$$\mathbf{R}_{\mathbf{x}}^{-1} = \frac{1}{\beta^2} \mathbf{J}_D. \tag{39b}$$

The representation of (38) is a minimum variance representation of a $1^{st}$ order GMr process $X$ whose covariance is $\beta^2 \mathbf{J}_D^{-1}$ [11]. The sinusoidal family of unitary transforms —including discrete cosine transform (DCT), discrete sine transform (DST), and their variations— is the class of complete orthonormal sets of eigenvectors generated by these Jacobi matrices $\mathbf{J}_D$ [11]. For example, with $k_1=k_2=1$, $k_3=k_4=0$, the eigenvectors of covariance matrices of $\mathbf{x}$ and $\mathbf{v}$ correspond to even DCT, $\mathbf{Q}_{DCT}$, defined as below:

$$[\mathbf{Q}_{DCT}]_{nk} = \begin{cases} \sqrt{\dfrac{1}{M}} & k=1, 1 \leq n \leq M \\ \sqrt{\dfrac{2}{M}} \cos \dfrac{\pi(2n-1)(k-1)}{2M}, & \begin{array}{l} 2 \leq k \leq M \\ 1 \leq n \leq M \end{array} \end{cases}. \tag{40}$$

This sinusoidal family of unitary transforms is also called as fast transforms because they can be implemented using fast algorithms such as DFT algorithm. In fact, fast sinusoidal transforms are Karhunen-Loeve (KL) transforms of non-stationary Markov processes [11].

In the next section, we present our results on the eigenvectors of the covariance matrix of $1^{st}$ and $2^{nd}$ order dfBm processes.

### III. STUDY OF THE EIGENVECTORS OF THE COVARIANCE MATRIX OF M-DFBM

In this section, we study the eigenvectors of the covariance matrix of a vector of discrete-time fractional Brownian motion (dfBm) process. Consider a length $M$ vector $\mathbf{x}(n)$ of random variables of a zero mean $m^{th}$ order dfBm process. The auto-covariance matrix, $\mathbf{R}_{B,m}^H(n)$, for any $n$ is a real symmetric and hence, diagonalizable matrix. Let the spectral decomposition of $\mathbf{R}_{B,m}^H(n)$ be given as

$\mathbf{R}_{B,m}^{H}(n) = \mathbf{Q}_{B,m}^{H}(n)\mathbf{\Lambda}_{B,m}^{H}(n)\mathbf{Q}_{B,m}^{H}{}^{T}(n)$, where $\mathbf{Q}_{B,m}^{H}(n)$ is an orthogonal matrix and $\mathbf{\Lambda}_{B,m}^{H}(n)$ is a diagonal matrix. We now present our results on the eigenvectors of the covariance matrix of the 1$^{st}$ and 2$^{nd}$ order dfBm using the analytic perturbation theory of linear operators.

*A. Eigenvectors of the Covariance Matrix of a 1-dfBm process with H=1/2*

Consider the *M*x*M* covariance matrix, $\mathbf{R}_{B,1}^{H=1/2}(n)$, of vector **x**(*n*) for a 1-dfBm process with *H*=0.5 (say for *M*=5). On using (7), we obtain

$$\mathbf{R_x}(n) = \mathbf{R}_{B,1}^{H=1/2}(n) = n\begin{bmatrix} 1 & 1 & 1 & 1 & 1 \\ 1 & 1+\frac{1}{n} & 1+\frac{1}{n} & 1+\frac{1}{n} & 1+\frac{1}{n} \\ 1 & 1+\frac{1}{n} & 1+\frac{2}{n} & 1+\frac{2}{n} & 1+\frac{2}{n} \\ 1 & 1+\frac{1}{n} & 1+\frac{2}{n} & 1+\frac{3}{n} & 1+\frac{3}{n} \\ 1 & 1+\frac{1}{n} & 1+\frac{2}{n} & 1+\frac{3}{n} & 1+\frac{4}{n} \end{bmatrix}, \quad (41)$$

where $\sigma_{H=1/2}^{2} = 1$, *B* in the subscript of $\mathbf{R}_{B,1}^{H=1/2}(n)$ implies that this auto-covariance matrix corresponds to a dfBm process and '1' in the subscript implies that this is a 1-dfBm process.

*Theorem 3.1:* The *M* x *M* auto-covariance matrix, $\mathbf{R}_{B,1}^{H=1/2}(n)$, of a vector, **x**(*n*), of random variables of a 1$^{st}$ order discrete-time fractional Brownian motion with *H*=1/2 is diagonalizable by the DCT matrix, i.e., the columns of DCT matrix serve as the eigenvectors of $\mathbf{R}_{B,1}^{H=1/2}(n)$ in the asymptotic sense.

*Proof:* It can be easily seen that for any *M* x *M* matrix $\mathbf{R}_{B,1}^{H=1/2}(n)$ in (41), the inverse is a tri-diagonal Jacobi matrix (equation (37)) with $\alpha = 0.5, k_1 = 1 - \frac{1}{n}, k_2 = 1, k_3 = k_4 = 0$ given as below:

$$\left[\mathbf{R}_{B,1}^{H=1/2}(n)\right]^{-1} = \frac{1}{2}\begin{bmatrix} 1-0.5\left(1-\frac{1}{n}\right) & -0.5 & 0 & . & . & . & 0 \\ -0.5 & 1 & -0.5 & 0 & . & . & 0 \\ 0 & -0.5 & 1 & -0.5 & 0 & . & . \\ 0 & 0 & . & . & . & . & . \\ . & . & . & . & . & . & 0 \\ . & . & . & . & . & 1 & -0.5 \\ 0. & . & . & 0 & 0 & -0.5 & 1-0.5 \end{bmatrix} = \frac{1}{\beta^2}\mathbf{J}_D(1-1/n,1,0,0) \quad (42)$$

On comparing (42) with (39b), we can see clearly that (42) corresponds to a vector of non-causal nonstationary GMr process of order 1. In fact, 1-dfBm with $H=1/2$ corresponds to the Wiener process that is indeed a GMr process of order one [12].

Further, from (42), it is observed that as $n \to \infty$, $\mathbf{J}_D(1-1/n,1,0,0) \to \mathbf{J}_D(1,1,0,0)$. As discussed in Section 2D above, $\mathbf{J}_D(1,1,0,0)$ is diagonalizable by $\mathbf{Q}_{DCT}$. Thus, the columns of $\mathbf{Q}_{DCT}$ are the eigenvectors of the auto-covariance matrix of 1-dfBm with $H=1/2$ in the asymptotic sense. ∎

*Lemma 3.1:* The $M \times M$ auto-covariance matrix, $\mathbf{R}_{B,1}^{H=1/2}(n)$, of a vector, $\mathbf{x}(n)$, of random variables of a $1^{st}$ order discrete-time fractional Brownian motion with $H=1/2$ has two distinct eigenvalues, $\lambda_1^{H=1/2} = 0$ with multiplicity $(M-1)$ and $\lambda_M^{H=1/2}(n) = nM$ with multiplicity one.

*Proof:* From (41), it is observed that as $n$ increases,

$$\mathbf{R}_{B,1}^{H=1/2}(n)\Big|_{\text{large } n} = n \begin{bmatrix} 1 & 1 & . & . & . & 1 \\ . & . & . & . & . & . \\ . & . & 1 & . & . & . \\ . & . & . & 1 & . & . \\ . & . & . & . & . & . \\ 1 & . & . & . & . & 1 \end{bmatrix} = n\mathbf{A} \quad (43)$$

where $\mathbf{A}$ is a singular matrix with two distinct eigenvalues, $\lambda_1 = 0$ with multiplicity $(M-1)$ (i.e., $\lambda_1 = \lambda_2 = ... = \lambda_{M-1} = 0$) and $\lambda_M = M$ with multiplicity one. Thus, a 1-dfBm process vector with $H=1/2$ has two distinct eigenvalues, $\lambda_1^{H=1/2} \stackrel{\Delta}{=} \lambda_1 = 0$ and $\lambda_M^{H=1/2}(n) \stackrel{\Delta}{=} \lambda_M = nM$. ∎

*B. Eigenvectors of the Covariance Matrix of a 1-dfBm process with $0<H<1$*

Consider the $M \times M$ auto-covariance matrix, $\mathbf{R}_{B,1}^H(n)$, of a vector $\mathbf{x}(n)$ for 1-dfBm process with $0<H<1$. In order to understand the eigenvectors of $\mathbf{R}_{B,1}^H(n)$, we rewrite the approximation (13) of $\mathbf{R}_{B,1}^H(n)$ using the below stated theorem.

*Theorem 3.2:* The $M \times M$ auto-covariance matrix, $\mathbf{R}_{B,1}^{H}(n)$, of a vector, $\mathbf{x}(n)$, of random variables of a 1st order discrete-time fractional Brownian motion with $0<H<1$ can be approximated as (here we use the symbol '^' to denote approximation)

$$\hat{\mathbf{R}}_{B,1}^{H}(n) = \frac{\sigma_H^2}{2} 2n^{2H} \left( \mathbf{A}_{H=1/2}(n) + \varepsilon_{1,1}^{H}(H\mathbf{B} - \mathbf{D}) + \varepsilon_{2,1}^{H}\mathbf{C} \right), \qquad \text{for } n > n_{\min} \qquad (44)$$

where

$$\mathbf{A}_{H=1/2}(n) = \frac{1}{n}\mathbf{R}_{B,1}^{H=1/2}(n) = \begin{bmatrix} 1 & 1 & 1 & 1 & & 1 \\ 1 & 1+\frac{1}{n} & . & . & & 1+\frac{1}{n} \\ 1 & . & 1+\frac{2}{n} & . & & 1+\frac{2}{n} \\ 1 & . & . & . & & . \\ 1 & 1+\frac{1}{n} & 1+\frac{2}{n} & . & & 1+\frac{(M-1)}{n} \end{bmatrix}, \quad \mathbf{B} = \begin{bmatrix} 0 & 1 & 2 & . . & (M-1) \\ 1 & 2 & 3 & . . & M \\ 2 & 3 & . & . . & (M+1) \\ . & . & . & . . & . \\ (M-2) & . & . & . . & (2M-3) \\ (M-1) & . & . & . . & (2M-2) \end{bmatrix},$$

$$\mathbf{C} = -\begin{bmatrix} 0 & 1 & 2^{2H} & . . & (M-1)^{2H} \\ 1 & 0 & 1 & . . & (M-2)^{2H} \\ . & . & . & . . & . \\ . & . & . & . . & . \\ . & . & . & . . & 1 \\ (M-1)^{2H} & . & . & . . & 0 \end{bmatrix}, \quad \mathbf{D} = \begin{bmatrix} 0 & 0 & 0 & . & . & 0 \\ 0 & 1 & 1 & . & . & 1 \\ . & 1 & 2 & . & . & 2 \\ . & . & . & . & . & . \\ . & . & . & . & (M-2) & (M-2) \\ 0 & 1 & 2 & . & (M-2) & (M-1) \end{bmatrix}, \varepsilon_{1,1}^{H} = \frac{1}{n}, \text{ and } \varepsilon_{2,1}^{H} = \frac{1}{2n^{2H}}.$$

The normalized approximation error, $e_{\hat{\mathbf{R}}}(n)$, defined by using the Frobenius norm of the difference between the true $\mathbf{R}_{B,1}^{H}(n)$ and its approximation $\hat{\mathbf{R}}_{B,1}^{H}(n)$ is bounded from above as

$$e_{\hat{\mathbf{R}}}(n) \stackrel{\Delta}{=} \frac{\| \mathbf{R}_{B,1}^{H}(n) - \hat{\mathbf{R}}_{B,1}^{H}(n) \|_F}{\| \mathbf{R}_{B,1}^{H}(n) \|_F} \leq |\alpha|, \qquad (45)$$

beyond $n > n_{\min}$ with $n_{\min} = \left\lceil (M-1)H \cdot \left( \left|2 - \frac{1}{H}\right| \cdot \frac{1}{|\alpha|} \right)^{\frac{1}{2}} \right\rceil$.

*Proof:* This is to note that the matrix $\mathbf{A}$ defined in (13) is the same as

$$\mathbf{A} = \mathbf{A}_{H=1/2}(n) - \varepsilon_{1,1}^{H}\mathbf{D} \qquad (46)$$

of (44) in Theorem 3.2 above. This shows that the approximation $\hat{\mathbf{R}}_{B,1}^{H}(n)$ defined in (13) and (44) are identical. Thus, the above theorem stands proved on the lines identical to the proof of Theorem 2.1 (refer to [7]). ∎

*Lemma 3.2:* The *M* x *M* auto-covariance matrix, $\mathbf{R}_{B,1}^{H}(n)$, of a vector, **x**(*n*), of random variables of a 1st order discrete-time fractional Brownian motion with 0<*H*<1 can be modeled as a linear perturbation of $\mathbf{R}_{B,1}^{H=1/2}(n)$ where the elements of $\mathbf{R}_{B,1}^{H}(n)$ are analytic functions of $\varepsilon$ in a neighborhood of $\varepsilon = 0$ such that

$$\hat{\mathbf{R}}_{B,1}^{H}(n) = \mathbf{R}_{B,1}^{H=1/2}(n) + \varepsilon \mathbf{R}_{B,1}^{H=1/2\,(1)}(n) \tag{47}$$

where $\varepsilon$ is a real number.

*Proof:* From Theorem 3.2, we can write the *M* x *M* auto-covariance matrix, $\mathbf{R}_{B,1}^{H}(n)$ as

$$\hat{\mathbf{R}}_{B,1}^{H}(n) = \frac{\sigma_H^2}{2} 2n^{2H}\left(\mathbf{A}_{H=1/2}(n) + \varepsilon_{1,1}^{H}(H\mathbf{B} - \mathbf{D}) + \varepsilon_{2,1}^{H}\mathbf{C}\right), \qquad \text{for } n > n_{\min} \tag{48}$$

This is to note that $\varepsilon_{1,1}^{H}$ and $\varepsilon_{2,1}^{H}$ defined in Theorem 3.2 are small and tend to zero as time index *n* increases beyond $n_{\min}$. The quantity

$$\left(\varepsilon_{1,1}^{H}(H\mathbf{B} - \mathbf{D}) + \varepsilon_{2,1}^{H}\mathbf{C}\right) = \varepsilon\left(\frac{\varepsilon_{1,1}^{H}}{\varepsilon_{2,1}^{H}}(H\mathbf{B} - \mathbf{D}) + \mathbf{C}\right)$$

$$= \varepsilon \mathbf{A}_{H=1/2}^{(1)}(n) \tag{49}$$

with $\varepsilon \equiv \varepsilon_{2,1}^{H}$ can be considered as a small linear perturbation in matrix $\mathbf{A}_{H=1/2}(n)$. Or, in other words, we can rewrite (47) as

$$\hat{\mathbf{R}}_{B,1}^{H}(n) = \frac{\sigma_H^2}{2} 2n^{2H}\left(\mathbf{A}_{H=1/2}(n) + \varepsilon \mathbf{A}_{H=1/2}^{(1)}(n)\right) \tag{50a}$$

$$= \mathbf{R}_{B,1}^{H=1/2}(n) + \varepsilon \mathbf{R}_{B,1}^{H=1/2\,(1)}(n) \tag{50b}$$

In order for the perturbation analysis to be valid, we choose *n* to be larger than $n_{\min}$ such that $\varepsilon_{1,1}^{H}$ and $\varepsilon_{2,1}^{H}$ (that are constant quantities for particular value of *n*) lie in the $|\varepsilon|$ radius of the neighborhood of zero. ∎

In Lemma 3.1, we have shown that as *n* increases $\mathbf{A}_{H=1/2}(n) \to \mathbf{A}$ and hence, $\mathbf{R}_{B,1}^{H=1/2}(n) \to n\mathbf{A}$. Thus, we can further simplify (50a) and write

$$\hat{\mathbf{R}}_{B,1}^{H}(n) = \frac{\sigma_H^2}{2} 2n^{2H}\left(\mathbf{A}_{H=1/2}(n) + \varepsilon \mathbf{A}_{H=1/2}^{(1)}(n)\right)$$

$$= \frac{\sigma_H^2}{2} 2n^{2H} \left( \mathbf{A} + \varepsilon \mathbf{A}^{(1)} \right) \tag{51a}$$

$$\frac{2\hat{\mathbf{R}}_{B,1}^H(n)}{2n^{2H}\sigma_H^2} = \mathbf{A} + \varepsilon \mathbf{A}^{(1)} \tag{52b}$$

Or, $\quad\quad \mathbf{A}(\varepsilon) = \mathbf{A} + \varepsilon \mathbf{A}^{(1)} \tag{52c}$

where $\mathbf{A}(\varepsilon) = \frac{2\hat{\mathbf{R}}_{B,1}^H(n)}{2n^{2H}\sigma_H^2}$. Equation (51) will help us study the eigenvectors of $\hat{\mathbf{R}}_{B,1}^H(n)$. From Lemma 3.1, we can easily see that the resolvent of $\mathbf{A}$ is

$$\mathbf{R}_z = (z\mathbf{I} - \mathbf{A})^{-1} = \sum_{r=0}^{M-1} \frac{r!}{z^{r+1}} \mathbf{Z}_{1r} + \frac{\mathbf{Z}_{20}}{\left(z - \lambda_M^{H=1/2}\right)} \tag{52}$$

where $\mathbf{Z}_{10}$ and $\mathbf{Z}_{20}$ are the projection matrices corresponding to the eigenvalues $\lambda_1^{H=1/2}$ and $\lambda_M^{H=1/2}$, respectively. The component matrices $\mathbf{Z}_{1r}$ for $r = 1,\ldots, (M\text{-}1)$ corresponding to $\lambda_1^{H=1/2}$ can be determined using (19).

Because $\mathbf{A}(\varepsilon)$ is a symmetric, toeplitz, and nonsingular matrix, the left and right eigenvectors of this matrix are identical. On using the perturbed matrix $\mathbf{A}(\varepsilon)$ from (51), we observe that the eigenvectors $\boldsymbol{\varphi}_j^H(\varepsilon) = \boldsymbol{\varphi}_j^{H=1/2} + \varepsilon \boldsymbol{\varphi}_j^{(1)} + \varepsilon^2 \boldsymbol{\varphi}_j^{(2)} + ..$ of $\mathbf{A}(\varepsilon)$ are the eigenvectors of $\hat{\mathbf{R}}_{B,1}^H(n)$. In order to model the eigenvalues and the eigenvectors of $\hat{\mathbf{R}}_{B,1}^H(n)$, we use result-2.2 from Section 2C above [10]. The first eigenvalue $\lambda_1^{H=1/2} = 0$ of $\mathbf{A}$ splits into $M$-1 eigenvalues $\lambda_{11}^H(\varepsilon), \lambda_{12}^H(\varepsilon), \ldots, \lambda_{1(M-1)}^H(\varepsilon)$ for sufficiently small $|\varepsilon|$ such that $\lambda_{11}^H(0) = \lambda_1^{H=1/2} = 0$ (for $r=1,2,\ldots M$-1). First $M$-1 eigenvalues of $\hat{\mathbf{R}}_{B,1}^H(n)$, denoted as $\lambda_{1r}^H(\varepsilon)$, are modeled as [7]

$$\lambda_{1r}^H(\varepsilon) = \frac{\sigma_H^2}{2} 2n^{2H} \left[ \lambda_1^{H=1/2} + \varepsilon a_{1r}^H + 0(\varepsilon)^{1+(1/l)} \right]$$

$$\approx \frac{\sigma_H^2}{2} a_{1r}^H \quad \text{for large } n \quad r=1,2,\ldots M\text{-}1 \tag{53}$$

where $a_{1r}^H$ are the eigenvalues of $\mathbf{Z}_{10}\mathbf{A}^{(1)}\mathbf{Z}_{10}$, and $\mathbf{C}$ is a constant matrix independent of time index $n$ and is defined in Theorems 2.1 and 3.2 above. Note that the first $M$-1 eigenvalues of $\hat{\mathbf{R}}_{B,1}^H(n)$ are given by (53). For the sake of notational convenience, we define

$$\lambda_r^H(\varepsilon) \stackrel{\Delta}{=} \lambda_{1r}^H(\varepsilon) \qquad r=1,2,...M\text{-}1 \tag{54}$$

The last eigenvalue of $\hat{\mathbf{R}}_{B,1}^H(n)$ was modeled in [7] as

$$\lambda_M^H(n) = \frac{\sigma_H^2}{2}\left[2Mn^{2H} + 2Hn^{2H-1}M(M-1)\right] + \frac{\sigma_H^2}{2}\left[-\frac{2}{M}\sum_{i=1}^{M-1}i(M-i)^{2H} + 2H^2n^{2H-2}\frac{M(M^2-1)}{12}\right]. \tag{55}$$

*Lemma 3.3:* All the eigenvectors of $\hat{\mathbf{R}}_{B,1}^H(n)$ are time-invariant in the asymptotic sense.

*Proof:* Corresponding to each $\lambda_{1r}^H(\varepsilon)$ in (53), there is an eigenvector $\boldsymbol{\varphi}_{1r}^H(\varepsilon)$ such that, for small enough $|\varepsilon|$

$$\boldsymbol{\varphi}_{1r}^H(\varepsilon) = \boldsymbol{\varphi}_r^H + \varepsilon^{1/l}\boldsymbol{\varphi}_{1,1} + \varepsilon^{2/l}\boldsymbol{\varphi}_{1,2} + ...,$$

$$\approx \boldsymbol{\varphi}_r^H \quad \text{for large } n \qquad r=1,2,....M\text{-}1, \tag{56}$$

where $\quad \boldsymbol{\varphi}_r^H \in Ker\left(\mathbf{A} - \lambda_1^{H=1/2}\mathbf{I}\right),$ (57)

and $\quad a_{1r}^H\boldsymbol{\varphi}_r^H = \mathbf{Z}_{10}\mathbf{A}^{(1)}\mathbf{Z}_{10}\boldsymbol{\varphi}_r^H.$ (58)

Note that these $M$-1 eigenvectors in (56) are the first $M$-1 eigenvectors of $\hat{\mathbf{R}}_{B,1}^H(n)$. For the sake of notational convenience, we define

$$\boldsymbol{\varphi}_r^H(\varepsilon) \stackrel{\Delta}{=} \boldsymbol{\varphi}_{1r}^H(\varepsilon) \qquad r=1,2,...M\text{-}1 \tag{59}$$

as the first $M$-1 eigenvectors of $\hat{\mathbf{R}}_{B,1}^H(n)$.

In order to understand the structure of $\boldsymbol{\varphi}_r^H(\varepsilon)$, we look at the structure of $\mathbf{Z}_{10}\mathbf{A}^{(1)}\mathbf{Z}_{10}$. On using (19) and (20), we can easily verify that

$$\mathbf{Z}_{10} = \mathbf{I} - \frac{1}{M}\mathbf{A} \tag{60a}$$

$$\mathbf{Z}_{1r} = \mathbf{A}\mathbf{Z}_{1r-1} \qquad r=1,2,...M\text{-}2 \tag{60b}$$

$$\mathbf{Z}_{20} = \frac{1}{M}\mathbf{A} \tag{60c}$$

$$\mathbf{E}_1 = -\frac{1}{M}\mathbf{Z}_{20} \tag{60d}$$

$$\mathbf{E}_2 = \frac{1}{M}\mathbf{Z}_{10} \tag{60e}$$

On using (51a) and (60a), we obtain

$$\mathbf{Z}_{10}\mathbf{A}^{(1)}\mathbf{Z}_{10} = \mathbf{Z}_{10}\left(\frac{\varepsilon_1}{\varepsilon}\mathbf{B} + \mathbf{C}\right)\mathbf{Z}_{10} = \mathbf{Z}_{10}\mathbf{C}\mathbf{Z}_{10} \quad (\text{because } \mathbf{Z}_{10}\mathbf{B}\mathbf{Z}_{10} = 0) \tag{61}$$

On using (61) and (58), we obtain

$$a_{1r}^H \boldsymbol{\varphi}_r^H = \mathbf{Z}_{10}\mathbf{A}^{(1)}\mathbf{Z}_{10}\boldsymbol{\varphi}_r^H$$

$$= \mathbf{Z}_{10}\mathbf{C}\mathbf{Z}_{10}\boldsymbol{\varphi}_r^H \tag{62}$$

where **C** is a constant matrix that depends on only the Hurst exponent $H$ and is independent of the time index $n$. Further,

$$\text{rank}(\mathbf{Z}_{10}\mathbf{C}\mathbf{Z}_{10}) = M\text{-}1 \tag{63}$$

Thus, we see that the first $M$-1 eigenvectors of $\hat{\mathbf{R}}_{B,1}^H(n)$ are the eigenvectors $\boldsymbol{\varphi}_r^H$ of $\mathbf{Z}_{10}\mathbf{C}\mathbf{Z}_{10}$ corresponding to its non-zero eigenvalues $a_{1r}^H$.

*Inference 1:* From (56), (62), and (63), we conclude that the first $M$-1 eigenvectors of $\hat{\mathbf{R}}_{B,1}^H(n)$ vary with only $H$ and are time-invariant in the asymptotic sense.

The last eigenvector of $\hat{\mathbf{R}}_{B,1}^H(n)$ can be modeled as

$$\boldsymbol{\varphi}_M^H(\varepsilon) = \boldsymbol{\varphi}_M^{H=1/2} + \varepsilon\boldsymbol{\varphi}_M^{(1)} \tag{64}$$

where $\boldsymbol{\varphi}_M^{H=1/2} = \frac{1}{\sqrt{M}}[1 \quad . \quad . \quad 1 \quad 1]^T$ and the perturbation $\boldsymbol{\varphi}_M^{(1)}$ can be computed using (27) as

$$\boldsymbol{\varphi}_M^{(1)} = \mathbf{E}_2\mathbf{A}^{(1)}\boldsymbol{\varphi}_M^{H=1/2} \tag{65}$$

On using (60e) in (65), we obtain

$$\boldsymbol{\varphi}_M^{(1)} = \frac{1}{M} \mathbf{Z}_{10} \left( \frac{\varepsilon_{1,1}^H}{\varepsilon_{2,1}^H} H\mathbf{B} + \mathbf{C} \right) \boldsymbol{\varphi}_M^{H=1/2} \tag{66}$$

where the two terms on the R.H.S. of (66) can be computed as

$$\frac{\varepsilon_{1,1}^H}{\varepsilon_{2,1}^H} H \frac{1}{M} \mathbf{Z}_{10} \mathbf{B} \boldsymbol{\varphi}_M = \left( \frac{\varepsilon_{1,1}^H}{\varepsilon_{2,1}^H} H \right) M(M-1) \boldsymbol{\varphi}_M^{H=1/2} \tag{67}$$

$$\frac{1}{M} \mathbf{Z}_{10} \mathbf{C} \boldsymbol{\varphi}_M = \left( -\frac{2}{M} \sum_{i=1}^{M-1} i.(M-i)^{2H} \right) \boldsymbol{\varphi}_M^{H=1/2} \tag{68}$$

On substituting (66) in (64), we obtain

$$\boldsymbol{\varphi}_M^H(\varepsilon) = \boldsymbol{\varphi}_M^{H=1/2} + \varepsilon \boldsymbol{\varphi}_M^{(1)}$$

$$= \left( 1 + \frac{HM(M-1)}{n} - \frac{1}{n^{2H} M} \sum_{i=1}^{M-1} i.(M-i)^{2H} \right) \boldsymbol{\varphi}_M^{H=1/2} \tag{69}$$

From (69), it is evident that the perturbation in the last eigenvector decreases rapidly with increase in *n*.

*Inference 2:* From above, the perturbed eigenvector $\boldsymbol{\varphi}_M^H(\varepsilon)$ can be considered to be time-invariant in the asymptotic sense.

From Inferences 1 and 2, Lemma 3.3 stands proved. ∎

From equations (56) to (59) and (69), we define a matrix consisting of the perturbed eigenvectors as below:

$$\hat{\mathbf{Q}}_{B,1}^H(n) \stackrel{\Delta}{=} \left[ \boldsymbol{\varphi}_1^H(\varepsilon) \quad . \quad . \quad . \quad \boldsymbol{\varphi}_{M-1}^H(\varepsilon) \quad \boldsymbol{\varphi}_M^H(\varepsilon) \right] \tag{70}$$

such that $\hat{\mathbf{Q}}_{B,1}^H(n)$ diagonalizes $\hat{\mathbf{R}}_{B,1}^H(n)$ in the asymptotic sense and also approaches a constant matrix as *n* increases (Lemma 3.3).

*Theorem 3.3:* The *M* x *M* auto-covariance matrix $\mathbf{R}_{B,1}^H(n)$ of a discrete-time 1$^{st}$ order fractional Brownian motion, with 0<*H*<1 is diagonalizable by the DCT matrix in the asymptotic sense. Or, in other words, $\mathbf{R}_{B,1}^H(n)$ can be approximated as $\widetilde{\mathbf{R}}_{B,1}^H(n)$ where

$$\widetilde{\mathbf{R}}_{B,1}^{H}(n) = \mathbf{Q}_{DCT}\,\widetilde{\mathbf{\Lambda}}_{B,1}^{H}(n)\mathbf{Q}_{DCT}^{T}, \tag{71}$$

for large $n$ in the asymptotic sense.

*Proof:* For the case of large $n$, the perturbation is very small. In that case, the proof of this theorem follows easily from Theorem 3.1, Lemma 3.2, and Lemma 3.3. ∎

In order to obtain the above approximation, we suggest Algorithm-A.

*Algorithm-A:*

1. Choose values of $H$ and $M$.

2. Form an $M \times M$ auto-covariance matrix, $\mathbf{R}_{B,1}^{H}(n)$, of a 1-dfBm process for a particular value of $n > n_{\min}$.

3. Carry out the eigenvalue decomposition of this matrix using the DCT matrix $\mathbf{Q}_{DCT}$:

$$\mathbf{D}_{Q}(n) = \mathbf{Q}_{DCT}^{T}\mathbf{R}_{B,1}^{H}(n)\mathbf{Q}_{DCT} \tag{72}$$

4. Keep only the diagonal values of $\mathbf{D}_{Q}(n)$ and define the diagonal matrix $\widetilde{\mathbf{\Lambda}}_{B,1}^{H}(n)$ as given in (73).

$$\widetilde{\mathbf{\Lambda}}_{B,1}^{H}(n) \stackrel{\Delta}{=} diag\{\mathbf{D}_{Q}(n)\} \tag{73}$$

5. Form an approximation, $\widetilde{\mathbf{R}}_{B,m}^{H}(n)$, of $\mathbf{R}(n)$ using the DCT matrix and the diagonal matrix $\widetilde{\mathbf{\Lambda}}_{B,1}^{H}(n)$ given as below:

$$\widetilde{\mathbf{R}}_{B,1}^{H}(n) = \mathbf{Q}_{DCT}\,\widetilde{\mathbf{\Lambda}}_{B,1}^{H}(n)\mathbf{Q}_{DCT}^{T} \tag{74}$$

Here, $\widetilde{\mathbf{R}}_{B,m}^{H}(n)$ is the estimate of $\mathbf{R}_{B,m}^{H}(n)$ and is the component of $\mathbf{R}_{B,m}^{H}(n)$ diagonalizable by the DCT matrix. In order to assess the *goodness of the estimate* $\widetilde{\mathbf{R}}_{B,m}^{H}(n)$, we define the normalized approximation error $e_{\widetilde{\mathbf{R}}}(n)$ using the Frobenius norm of the difference between $\mathbf{R}_{B,m}^{H}(n)$ and its approximation $\widetilde{\mathbf{R}}_{B,m}^{H}(n)$ as:

$$e_{\widetilde{\mathbf{R}}}(n) \stackrel{\Delta}{=} \frac{\left\|\mathbf{R}_{B,m}^{H}(n) - \widetilde{\mathbf{R}}_{B,m}^{H}(n)\right\|_{F}}{\left\|\mathbf{R}_{B,m}^{H}(n)\right\|_{F}} \times 100\,\% \tag{75}$$

Similarly, we define the normalized approximation error $e_{\tilde{\Lambda}}(n)$ using the Frobenius norm of the difference between the eigenvalues of $\mathbf{R}_{B,m}^H(n)$ and of approximation $\tilde{\mathbf{R}}_{B,m}^H(n)$ given as below:

$$e_{\tilde{\Lambda}}(n) \stackrel{\Delta}{=} \frac{\left\| \Lambda_{B,m}^H(n) - \tilde{\Lambda}_{B,m}^H(n) \right\|_F}{\left\| \Lambda_{B,m}^H(n) \right\|_F} \times 100\% \qquad (76)$$

Simulation results to validate the above results are presented in Section 4.

*C. Eigenvectors of the Covariance Matrix of a 2-dfBm process*

In this subsection, we will look into the structure of a 2$^{nd}$ order dfBm. Consider a length $M$ vector, $\mathbf{x}(n)$, of random variables of a 2-dfBm. The auto-covariance sequence of this process can be computed using (8) given as below:

$$r_{B,2}^H(n_1, n_2) = \frac{\sigma_H^2}{2}\left(|n_1 - n_2|^{2H} - |n_1|^{2H} - |n_2|^{2H}\right) + \frac{\sigma_H^2}{2}\left(2Hn_1|n_2|^{2H-1} + 2Hn_2|n_1|^{2H-1}\right), \qquad (77)$$

where $1 < H < 2$.

On writing the $M \times M$ auto-covariance matrix of this process and simplifying the structure similar to Theorem 3.2, we notice that this covariance matrix can be approximated as:

$$\hat{\mathbf{R}}_{B,2}^H(n) = \frac{\sigma_H^2}{2} 2n^{2H}(2H-1)\left(\mathbf{A}_{H=1/2}(n) + \varepsilon_{1,2}^H(H\mathbf{B} - \mathbf{D}) - \varepsilon_{2,2}^H \mathbf{C}\right), \qquad \text{for } n > n_{\min} \qquad (78)$$

where 
$$\mathbf{A}_{H=1/2}(n) = \frac{1}{n}\mathbf{R}_{B,1}^{H=1/2}(n) = \begin{bmatrix} 1 & 1 & 1 & 1 & 1 \\ 1 & 1+\frac{1}{n} & . & . & 1+\frac{1}{n} \\ 1 & . & 1+\frac{2}{n} & . & 1+\frac{2}{n} \\ 1 & . & . & . & . \\ 1 & 1+\frac{1}{n} & 1+\frac{2}{n} & . & 1+\frac{(\dot{M}-1)}{n} \end{bmatrix}, \quad \mathbf{B} = \begin{bmatrix} 0 & 1 & 2 & .. & (M-1) \\ 1 & 2 & 3 & .. & M \\ 2 & 3 & . & .. & (M+1) \\ . & . & . & . & . \\ (M-2) & . & . & .. & (2M-3) \\ (M-1) & . & . & .. & (2M-2) \end{bmatrix},$$

$$\mathbf{C} = -\begin{bmatrix} 0 & 1 & 2^{2H} & .. & (M-1)^{2H} \\ 1 & 0 & 1 & .. & (M-2)^{2H} \\ . & . & . & .. & . \\ . & . & . & .. & . \\ . & . & . & .. & 1 \\ (M-1)^{2H} & . & . & .. & 0 \end{bmatrix}, \quad \mathbf{D} = \begin{bmatrix} 0 & 0 & 0 & . & . & 0 \\ 0 & 1 & 1 & . & . & 1 \\ . & 1 & 2 & . & . & 2 \\ . & . & . & . & . & . \\ . & . & . & . & (M-2) & (M-2) \\ 0 & 1 & 2 & . & (M-2) & (M-1) \end{bmatrix}, \varepsilon_{1,2}^H = \frac{1}{n},$$

and $\varepsilon_{2,2}^H = \dfrac{1}{2n^{2H}(2H-1)}$.

From (78), we observe that the structure of the auto-covariance matrix of a length M vector of 2$^{nd}$ order dfBm is similar to that of the 1$^{st}$ order dfBm. This shows that this matrix will also be diagonalizable by the DCT matrix in the asymptotic sense.

IV. SIMULATION RESULTS

From the theorems and results presented in Section 3 above, it is evident that the covariance matrix corresponding to the 1$^{st}$ and 2$^{nd}$ order dfBm processes are diagonalizable by the DCT matrix in the asymptotic sense. Or, in other words, we can say that $\mathbf{R}_{B,m}^H(n)$ for $m=1$ and 2 consists of two components where the larger component is diagonalizable by the DCT matrix. In this section, we would like to assess the validity of the work presented in Section 3. We conduct two experiments: 1) for a 1$^{st}$ order dfBm process and 2) for a 2$^{nd}$ order dfBm process.

*Experiment 1 (1$^{st}$ order dfBm):* In this experiment, our aim is to ascertain if the eigenvectors of $\mathbf{R}_{B,1}^H(n)$ are close to the DCT basis vectors. We follow the steps of Algorithm-B in this experiment.

*Algorithm-B:*

1. Follow Algorithm A to construct $\mathbf{R}_{B,1}^H(n)$ and obtain the estimates $\tilde{\mathbf{\Lambda}}_{B,1}^H(n)$ and $\tilde{\mathbf{R}}_{B,1}^H(n)$ (eq. (71)) using the DCT matrix for particular values of $n$, $M$, and $H$.

2. Carry out the spectral decomposition of $\mathbf{R}_{B,1}^H(n)$ given as $\mathbf{R}_{B,1}^H(n) = \mathbf{Q}_{B,1}^H(n)\mathbf{\Lambda}_{B,1}^H(n)\left(\mathbf{Q}_{B,1}^H(n)\right)^T$, where $\mathbf{Q}_{B,1}^H(n)$ is the unitary matrix that diagonalizes $\mathbf{R}_{B,1}^H(n)$ and $\mathbf{\Lambda}_{B,1}^H(n)$ is a diagonal matrix.

3. Using the analytical results in the previous section, calculate the perturbed matrix $\hat{\mathbf{Q}}_{B,1}^H$ (eq. 70) that diagonalizes $\hat{\mathbf{R}}_{B,1}^H(n)$ in the asymptotic sense.

4. Compute four normalized errors:

    a) $e_{\hat{\mathbf{Q}}}(n)$ between the actual diagonalizing matrix $\mathbf{Q}_{B,1}^H(n)$ and its approximation $\hat{\mathbf{Q}}_{B,1}^H(n)$ (Table 1),

b) $e_{\tilde{\mathbf{Q}}}(n)$ between the actual diagonalizing matrix $\mathbf{Q}_{B,1}^{H}(n)$ and the DCT matrix (Table 2),

c) $e_{\tilde{\Lambda}}(n)$ between the eigenvalues of $\mathbf{R}_{B,1}^{H}(n)$ and those of approximation $\tilde{\mathbf{R}}_{B,1}^{H}(n)$ (Table 3), and

d) $e_{\tilde{\mathbf{R}}}(n)$ between the actual auto-covariance matrix $\mathbf{R}_{B,1}^{H}(n)$ and its approximation $\tilde{\mathbf{R}}_{B,1}^{H}(n)$ (Table 4).

We present the above results for three values of $H$ (0.2, 0.45, and 0.8), two values of time index $n$ (200 and 2000), and for values of $M$ varying from 2 to 9. In addition to the above, we display results in Figures 1 to 3. Figure 1 displays error $e_{\tilde{\mathbf{R}}}(n)$ (measure of approximation by DCT matrix) versus $H$ for $M=5$ for three values of time index $n$ (200, 500, and 2000). Figures 2 and 3 display error $e_{\tilde{\mathbf{R}}}(n)$ versus $M$ for two values of $H$ (0.45 and 0.8), and three values of time index $n$ (200, 500, and 2000).

*Experiment 2 ($2^{nd}$ order dfBm):* In the second experiment, we compute the normalized error $e_{\tilde{\mathbf{R}}}(n)$ between the actual auto-covariance matrix $\mathbf{R}_{B,1}^{H}(n)$ and its approximation $\tilde{\mathbf{R}}_{B,1}^{H}(n)$ obtained from Algorithm A using the DCT matrix for $2^{nd}$ order dfBm. The results are tabulated in Table 5. In addition to the above, we plot Figures 4 and 5 that display error $e_{\tilde{\mathbf{R}}}(n)$ versus $M$ for two values of $H$ (1.2 and 1.55) and three values of time index $n$ (200, 500, and 2000).

TABLE-I
Normalized error $e_{\hat{\mathbf{Q}}}(n)$ (in %) between the actual diagonalizing matrix $\mathbf{Q}_{B,1}^{H}(n)$ and its approximation $\hat{\mathbf{Q}}_{B,1}^{H}(n)$

| M | n=200, H=0.2 | n=2000, H=0.2 | n=200, H=0.45 | n=2000, H=0.45 | n=200, H=0.8 | n=2000, H=0.8 |
|---|---|---|---|---|---|---|
| 2 | 0.0820 | 0.0080 | 0.1781 | 0.0178 | 0.3158 | 0.0316 |
| 3 | 0.1940 | 0.0713 | 0.1234 | 0.0126 | 0.2112 | 0.0211 |
| 4 | 0.5020 | 0.1880 | 0.1562 | 0.0160 | 0.3926 | 0.1098 |
| 5 | 0.5423 | 0.2016 | 0.3549 | 0.0366 | 0.4745 | 0.1346 |
| 6 | 0.2767 | 0.0988 | 0.2086 | 0.0214 | 0.5372 | 0.1515 |
| 7 | 0.5796 | 0.2119 | 0.2306 | 0.0236 | 0.5835 | 0.1728 |
| 8 | 0.2990 | 0.2139 | 0.2506 | 0.0257 | 0.8950 | 0.1817 |
| 9 | 0.5980 | 0.2150 | 0.4926 | 0.0509 | 0.6557 | 0.1754 |

TABLE-II
Normalized error $e_{\tilde{\mathbf{Q}}}(n)$ (in %) between the actual diagonalizing matrix $\mathbf{Q}_{B,1}^{H}(n)$ and the DCT matrix

| M | n=200, H=0.2 | n=2000, H=0.2 | n=200, H=0.45 | n=2000, H=0.45 | n=200, H=0.8 | n=2000, H=0.8 |
|---|---|---|---|---|---|---|
| 2 | 0.0531 | 0.0051 | 0.1127 | 0.0113 | 0.1995 | 0.0200 |
| 3 | 0.2727 | 0.1007 | 0.1662 | 0.0170 | 0.2825 | 0.0283 |
| 4 | 2.6047 | 2.5861 | 0.5137 | 0.4590 | 2.6653 | 2.8256 |
| 5 | 3.4915 | 3.4579 | 0.6691 | 0.6028 | 3.4596 | 3.6393 |
| 6 | 4.1735 | 4.1326 | 0.7930 | 0.7166 | 4.1227 | 4.3194 |
| 7 | 4.6706 | 4.6222 | 0.8853 | 0.7987 | 4.6057 | 4.8067 |
| 8 | 5.0760 | 5.0226 | 0.9622 | 0.8662 | 5.0069 | 5.2117 |
| 9 | 5.4022 | 5.3444 | 1.0253 | 0.9204 | 5.3307 | 5.5360 |

TABLE-III

Normalized error $e_{\tilde{\Lambda}}(n)$ (in %) between the actual eigenvalues of $\Lambda_{B,1}^H(n)$ and their approximation in $\tilde{\Lambda}_{B,1}^H(n)$

| M | n=200, H=0.2 | n=2000, H=0.2 | n=200, H=0.45 | n=2000, H=0.45 | n=200, H=0.8 | n=2000, H=0.8 |
|---|---|---|---|---|---|---|
| 2 | 3.8564 x10$^{-5}$ | 3.6638 x10$^{-7}$ | 0.0002 | 0.0179 x10$^{-4}$ | 0.0006 | 0.0057 x10$^{-3}$ |
| 3 | 0.0015 | 0.0002 | 0.0005 | 0.0519 x10$^{-4}$ | 0.0015 | 0.0151 x10$^{-3}$ |
| 4 | 0.0037 | 0.0011 | 0.0009 | 0.0926 x10$^{-4}$ | 0.0028 | 0.0277 x10$^{-3}$ |
| 5 | 0.0064 | 0.0021 | 0.0015 | 0.1467 x10$^{-4}$ | 0.0044 | 0.0440 x10$^{-3}$ |
| 6 | 0.0086 | 0.0029 | 0.0022 | 0.2130 x10$^{-4}$ | 0.0064 | 0.0639 x10$^{-3}$ |
| 7 | 0.0104 | 0.0035 | 0.0030 | 0.2919 x10$^{-4}$ | 0.0087 | 0.0874 x10$^{-3}$ |
| 8 | 0.0120 | 0.0041 | 0.0039 | 0.3835 x10$^{-4}$ | 0.0113 | 0.1145 x10$^{-3}$ |
| 9 | 0.0134 | 0.0046 | 0.0050 | 0.4879 x10$^{-4}$ | 0.0143 | 0.1454 x10$^{-3}$ |

TABLE-IV

Normalized error $e_{\tilde{R}}(n)$ (in %) between the actual auto-covariance matrix $R_{B,1}^H(n)$ and its approximation $\tilde{R}_{B,1}^H(n)$

| M | n=200, H=0.2 | n=2000, H=0.2 | n=200, H=0.45 | n=2000, H=0.45 | n=200, H=0.8 | n=2000, H=0.8 |
|---|---|---|---|---|---|---|
| 2 | 0.0727 | 0.0072 | 0.1590 | 0.0159 | 0.2821 | 0.0283 |
| 3 | 0.4610 | 0.1732 | 0.2720 | 0.0279 | 0.4596 | 0.0462 |
| 4 | 0.6428 | 0.2403 | 0.3768 | 0.0390 | 0.6279 | 0.0632 |
| 5 | 0.7754 | 0.2876 | 0.4780 | 0.0496 | 0.7923 | 0.0799 |
| 6 | 0.8837 | 0.3251 | 0.5773 | 0.0600 | 0.9546 | 0.0965 |
| 7 | 0.9775 | 0.3567 | 0.6754 | 0.0703 | 1.1153 | 0.1130 |
| 8 | 1.0617 | 0.3842 | 0.7725 | 0.0805 | 1.2747 | 0.1294 |
| 9 | 1.1391 | 0.4090 | 0.8689 | 0.0906 | 1.4331 | 0.1458 |

TABLE-V

Error $e_{\tilde{R}}(n)$ (in %) between the actual auto-covariance matrix $R_{B,2}^H(n)$ and its approximation $\tilde{R}_{B,2}^H(n)$ for a 2-dfBm process

| M | n=200, H=1.2 | n=2000, H=1.2 | n=200, H=1.45 | n=2000, H=1.45 | n=200, H=1.8 | n=2000, H=1.8 |
|---|---|---|---|---|---|---|
| 2 | 0.4232 | 0.0424 | 0.5114 | 0.0513 | 0.6348 | 0.0636 |
| 3 | 0.6894 | 0.0692 | 0.8330 | 0.0837 | 1.0340 | 0.1039 |
| 4 | 0.9416 | 0.0948 | 1.1377 | 0.1145 | 1.4123 | 0.1422 |
| 5 | 1.1881 | 0.1199 | 1.4355 | 0.1449 | 1.7819 | 0.1798 |
| 6 | 1.4312 | 0.1447 | 1.7292 | 0.1749 | 2.1463 | 0.2171 |
| 7 | 1.6718 | 0.1695 | 2.0200 | 0.2048 | 2.5071 | 0.2542 |
| 8 | 1.9106 | 0.1941 | 2.3083 | 0.2345 | 2.8648 | 0.2911 |
| 9 | 2.1476 | 0.2187 | 2.5947 | 0.2642 | 3.2200 | 0.3280 |

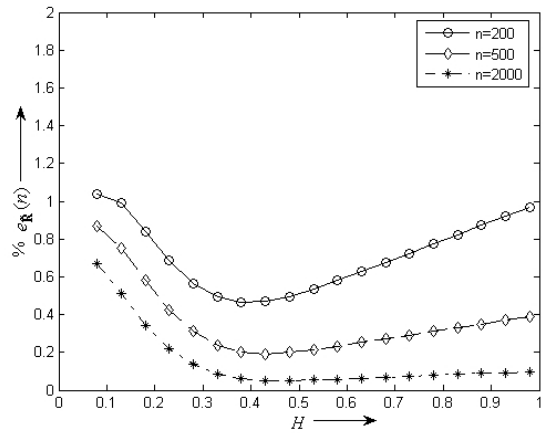

Figure 1: H vs. % $e_{\tilde{R}}(n)$ for three values of time index n for M=5 (1$^{st}$ order dfBm)

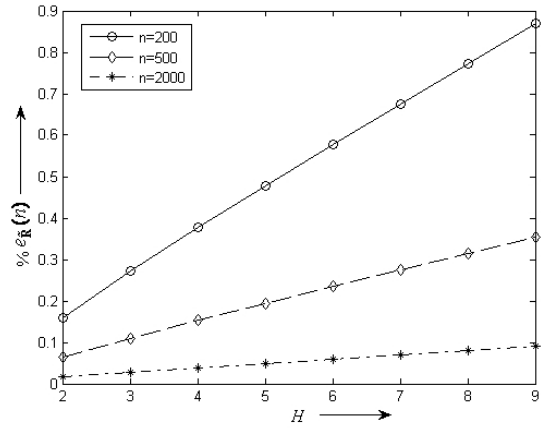

Figure 2: $M$ vs. % $e_{\widetilde{\mathbf{R}}}(n)$ corresponding to $H$=0.45 for three values of time index $n$ (1$^{st}$ order dfBm)

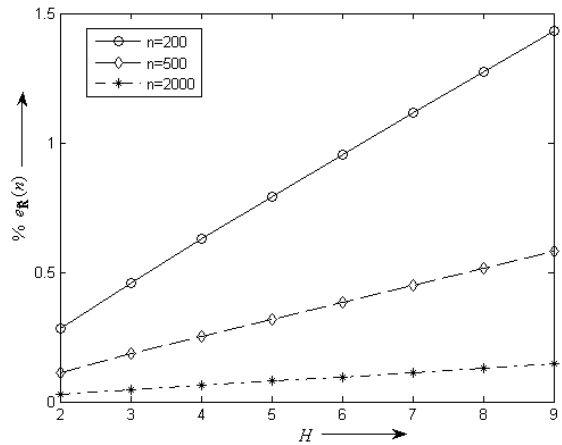

Figure 3: $M$ vs. % $e_{\widetilde{\mathbf{R}}}(n)$ corresponding to $H$=0.8 for three values of time index $n$ (1$^{st}$ order dfBm)

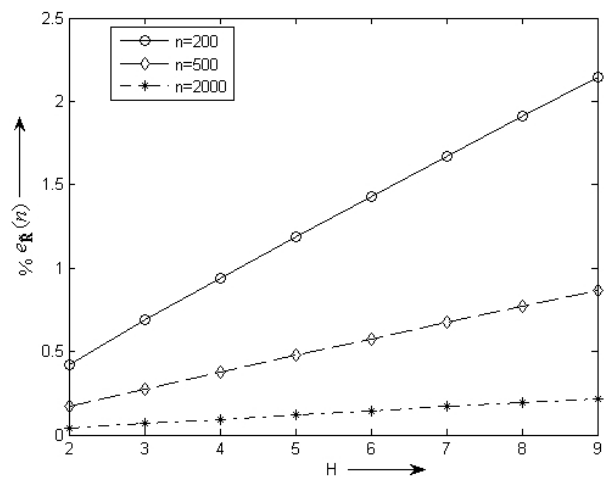

Figure 4: $M$ vs. % $e_{\widetilde{\mathbf{R}}}(n)$ corresponding to $H$=1.2 for three values of time index $n$ (2$^{nd}$ order dfBm)

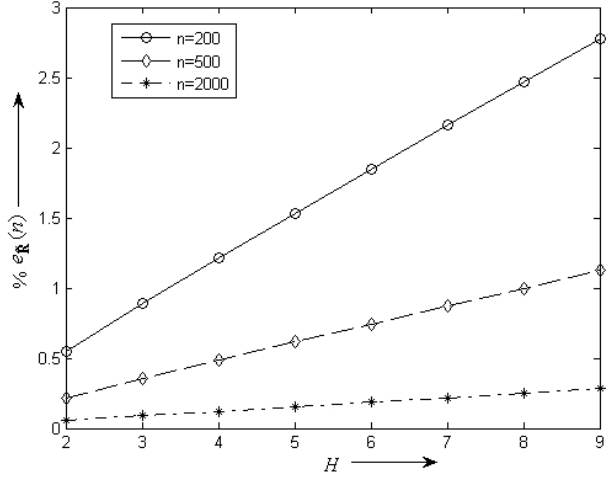

Figure 5: $M$ vs. % $e_{\widetilde{\mathbf{R}}}(n)$ corresponding to $H$=1.55 for three values of time index $n$ ($2^{\text{nd}}$ order dfBm)

The following observations are drawn from Tables 1 to 4 and Figures 1 to 3 for a 1-dfBm process:

1. From Table 1, we note that the normalized percentage error $e_{\hat{\mathbf{Q}}}(n)$ between the actual diagonalizing matrix $\mathbf{Q}_{B,1}^{H}(n)$ of $\mathbf{R}_{B,1}^{H}(n)$ and its approximation $\hat{\mathbf{Q}}_{B,1}^{H}(n)$ that is formed in (70) by modeling the perturbed eigenvectors in (59) and (69) is less than 1% for different values of $H$, $n$, or $M$. Further, as $n$ increases the error percentage decreases.

   *Discussion:* The low value of percentage error indicates that the analytical modeling of perturbed eigenvectors as presented in this paper is good. The decrease in error percentage with $n$ implies that the modeling error further reduces as $n$ increases and hence, the modeling is more accurate in the asymptotic sense.

2. From Table 2, we note that the normalized percentage error $e_{\widetilde{\mathbf{Q}}}(n)$ between the actual diagonalizing matrix $\mathbf{Q}_{B,1}^{H}(n)$ of $\mathbf{R}_{B,1}^{H}(n)$ and the DCT matrix is less than 6% for different values of $H$, $n$, or $M$. Further, it is noticed that as $M$ increases beyond 3, the error percentage does not change with the increase in $n$, but increases with $H$.

   *Discussion:* From the above, it is evident that the diagonalizing matrix $\mathbf{Q}_{B,1}^{H}(n)$ is a constant matrix independent of $n$ in the asymptotic sense. However, the eigenvectors are indeed a function of $H$.

3. The normalized percentage error norm, $e_{\tilde{\Lambda}}(n)$, between the norm of eigenvalues of $\mathbf{R}_{B,1}^{H}(n)$ and the eigenvalues obtained from the DCT diagonalizable approximation $\tilde{\mathbf{R}}(n)$ is less than 0.05% for any value of $H$ and $n > n_{\min}$.

   *Discussion:* This shows that for all practical purposes (such as subband decomposition) where we would like to work with the eigenvectors of the dfBm process [8], DCT basis can be conveniently used to yield equivalently good results.

4. From Table 4 and Figures 1 to 3, it is clear that for large $n$, the normalized percentage error $e_{\tilde{\mathbf{R}}}(n)$ using the DCT matrix approximation is minimum near $H=1/2$. Also, for large $n$, the percentage error corresponding to the range $0.5<H<1$ is less as compared to $0<H<0.5$. The same observation is noted for $e_{\tilde{\mathbf{Q}}}(n)$ from Table-2.

   *Discussion*: Note that the dfBm process with $H=1/2$ corresponds to a Wiener process that is a non-causal nonstationary GMr process of order 1 [12]. From Theorem 3.1, we notice that DCT is the diagonalizing transform for the covariance matrix of this process. Thus, the modeling error is least near $H=1/2$ and increases as we go beyond $H=1/2$ on either side.

*2-dfBm process:* From Table 5 and Figures 4 and 5, we note that for a 2-dfBm process, the normalized percentage error $e_{\tilde{\mathbf{R}}}(n)$ using the DCT matrix approximation $\mathbf{R}_{B,2}^{H}(n)$ is less than 3% for different values of $H$, $n$, or $M$. Again the percentage error decreases rapidly with increasing $n$ and reduces to less than 0.5%. Hence, similar to a 1-dfBm process, the covariance matrix of a 2-dfBm process can also be approximated using the DCT matrix.

## V.  CONCLUSIONS

In this paper, we established a connection between discrete cosine transform (DCT), and 1$^{st}$ and 2$^{nd}$ order discrete-time fractional Brownian motion (dfBm). It is shown that the 1$^{st}$ order dfBm with $H=1/2$ is a noncausal nonstationary Gauss Markov random process whose covariance matrix is diagonalizable by DCT matrix. It is further shown that the eigenvectors of the covariance matrix of any $1^{st}$ and $2^{nd}$ order dfBm can be approximated by DCT basis vectors in the asymptotic sense. The perturbation in the eigenvectors from DCT basis vectors corresponding to $H=1/2$ is modeled using the analytic perturbation theory of linear operators.